\documentstyle[12pt]{article}

\newcommand{\ket}[1]{\left| {#1} \right>}
\newcommand{\bet}[1]{\left|\rule{-0.2em}{1.5ex}\right.\overline{#1}
\left.\rule{-0.1em}{1.5ex}\right>}

\begin{document}

\title{Multiple Particle Interference and Quantum Error Correction} 

\author{Andrew Steane\thanks{Electronic 
address: a.steane@physics.oxford.ac.uk}\\
\small Clarendon Laboratory, Parks Road, Oxford, OX1 3PU, England.}

\date{November 20, 1995}
\maketitle

\begin{abstract}
The concept of {\em multiple particle interference} is discussed, using 
insights provided by the classical theory of error correcting codes. This 
leads to a discussion of error correction in a quantum communication 
channel or a quantum computer. Methods of error correction in the quantum 
regime are presented, and their limitations assessed. A quantum channel can 
recover from arbitrary decoherence of $x$ qubits if $K$ bits of quantum 
information are encoded using $n$ quantum bits, where $K/n$ can be greater 
than $1-2 H(2x/n)$, but must be less than $1 - 2 H(x/n)$. This
implies exponential reduction of decoherence with only a polynomial
increase in the computing resources required. Therefore
quantum computation can be made free of errors in the presence of 
physically realistic levels of decoherence.
The methods also allow isolation of quantum communication from noise
and evesdropping (quantum privacy amplification).
  \end{abstract}

\section{Introduction}

The concepts of quantum interference, correlations and entanglement are at 
the heart of quantum mechanics. A quantum interference between two parts
a system's evolution is prevented when the system interacts with another
so as to produce an entangled state. In such situations, properties
of the two entangled systems are correlated, and
correlations of this type are subject to the 
Bell inequalities (Bell 1964), which shows that they are non-local in 
character. Whereas for many quantum mechanical effects a model can be given 
which relies only on classical concepts, this non-locality is a feature of 
quantum mechanics which is alien to the very structure of other (classical) 
theories. (Many texts are available as an introduction to this broad 
subject, for example that of Shimony 1989.) 

One way of shedding light on the nature of quantum mechanics is to pose the 
question ``to what extent can quantum mechanical behaviour be modelled in 
classical terms?'' To make this slightly vague question more concrete, it 
can be posed thus: ``to what extent can quantum mechanical behaviour be 
simulated by means of a universal computer operating according to the laws 
of classical mechanics?'' Such a `universal' computer is universal in the 
sense of a universal Turing machine: it can simulate the behaviour of any 
other computer in the set of all possible computers (Turing 1936). However, 
as long as the set of `possible' computers includes only 
those operating by classical laws of physics, then the non-local 
correlations which arise in the real world cannot be simulated, as was 
discussed by Feynman 1982. To simulate them, the computer must be allowed to 
operate according to the laws of quantum mechanics. Hence one introduces 
the concept of the quantum, as opposed to classical, computer
(see Deutsch 1985; Ekert 1995), and the question under
consideration can be re-phrased: 
``to what extent can a quantum computer perform calculations which are 
beyond the computing abilities of a classical computer?'' For the 
physicist, this question addresses fundamental questions concerning the 
nature of our most basic physical theory. However, the answer is also of 
considerable practical interest because computing ability is an extremely 
useful kind of ability. 

The theoretical analysis of quantum computers has by now passed some 
important milestones, among them the demonstration of how to construct a 
universal quantum computer (Deutsch 1985), the discovery of simple 
universal quantum gates (Deutsch {\em et al.} 1995, Barenco 1995, 
DiVincenzo 1995), and the presentation of algorithms for idealised quantum 
computers which surpass the computing ability of known algorithms for 
classical computers, and which appear to surpass even what is in principle 
possible classically (Deutsch \& Jozsa 1992, Bernstein 1993, Shor 1994, 
Simon 1994). It has been obvious from the outset that quantum computation 
is different from classical computation precisely because of the 
possibility of quantum interference and entanglement. However, this 
entanglement is itself sensitive to a problem which is unavoidable in the 
quantum context, namely, {\em decoherence} of the state of the
computer (Zurek 1993, Landauer 1995). 
The useful algorithms just mentioned were initialy proposed under the 
assumption of the idealised case that this decoherence, or generally
any process involving loss of quantum information, is negligible. 
However, it can be argued that the possibility of decoherence is itself 
just as fundamental a feature of quantum mechanics as the interference and 
entanglement of which a quantum computer takes advantage. Such decoherence 
must be considered, for example, in any discussion of the ``Schr\"odinger's 
cat" paradox (Schr\"odinger 1935; for a text-book treatment see, for 
example, Peres 1993). The cat in Schr\"odinger's thought-experiment 
corresponds here to the quantum computer itself. Hence, the idealisation in 
which decoherence is taken to be negligible is not merely a limit on the 
practical application of the theory of quantum computation, it is in fact 
an ``idealisation too far'', since it involves neglecting a basic aspect of 
quantum theory, as has been emphasized by Landauer 1995. 

This paper discusses both the nature of quantum interference involving many 
particles, and also the question of decoherence in quantum computers. It is 
shown that both questions are intimately concerned with the issue of error 
correction which arises in classical information theory. Unruh 1995 and
Palma~{\em et al.}~1996
calculated the sensitivity of a `bare' quantum computer to thermal 
decoherence. Here, we are concerned with the different question of how to 
add redundancy to such a `bare' computer in order to stabilise it against 
all error processes, including decoherence. The classical theory of error correction which is invoked is a 
well-founded body of knowledge involving some beautiful mathematical 
concepts, and we can take advantage of this knowledge in our quest to 
understand quantum mechanics more fully. This paper does not assume much 
familiarity with classical error correction, however. At the risk of 
alienating experts, concepts from classical information theory are 
introduced for the most part with sufficient explanation to allow readers 
unfamiliar with this material to follow the argument. The readily available 
textbooks such as MacWilliams \& Sloane 1977 and Hamming 1986 give further 
explanation. 

In section \ref{sec:mpi} a general theory of interference involving many 
particles is presented. It is shown that an interesting class of 
entanglements involving many particles (or other simple quantum systems) 
can be understood by appealing to the known theory of classical error 
correcting codes. In section \ref{sec:ecq} the same concepts are applied to 
the problem of error correction in a set of two-state systems (`quantum 
bits'). Coding and correction methods are presented which allow the problem 
of decoherence in a quantum computer to be circumvented. The same methods 
allow privacy in quantum cryptography to be enhanced (for a review and 
references to this subject, see, for example, Hughes {\em et al.} 1995
and Phoenix and Townsend 1995.) In 
section \ref{sec:errlim} the limitations of these coding methods are 
estimated, by a calculation reminiscent of Shannon's main theorem for 
communication through a noisy channel. A full quantum equivalent to 
Shannon's theorem is not found, and this is a limitation of the present 
work, but the ideas presented here suggest ways of tackling this more 
general question. The conclusions of the present discussion are hopeful, 
however, in that they suggest that error-free quantum computation is 
possible using resources (numbers of quantum bits and of operations) that 
are only a polynomial factor greater than those required by an ideal 
quantum computer. Indeed, the judicious use of redundency and error correction
allows the probability of decoherence to {\em fall exponentially} with the
amount of redundency. This is a conclusion which has commonly been imagined
to be ruled out for quantum systems. The implementation of the error
correction procedure to be described, without introducing excessive
extra decoherence, remains a severe technological challenge, however.

\section{Multiple particle interference and parity checks} \label{sec:mpi}

\subsection{Single parity check}

Consider a two-state quantum system. Its two-dimensional Hilbert space is 
spanned by two orthogonal states which will be written $\ket{0}$ and 
$\ket{1}$. These states may for example be different states of motion of a 
spinless particle of no internal structure, or they may be different 
internal states, such as those of a two-level atom. The simple concept of 
quantum interference arises when such a system is in a state such as 
$(\ket{0} + \exp(i\phi)\ket{1})/\sqrt{2}$, and measurements are performed 
which project the state onto $(\ket{0} \pm \ket{1})/\sqrt{2}$. Now, what 
happens if this system interacts with another two-state system, such that 
the total state of the pair is the entangled state $(\ket{0}\otimes\ket{0} 
+ \exp(i\phi)\ket{1}\otimes\ket{1})/\sqrt{2}\,$? In this case, measurements 
on either subsystem alone (hereafter called a `particle') will not reveal 
any interference effect (any dependence on the value of $\phi$). If both 
particles are measured in the $(\ket{0} \pm \ket{1})/\sqrt{2}$ basis, on 
the other hand, and {\em the results of the measurements on each particle 
pair are compared}, then a correlation is observed which is sensitive to 
$\phi$. The probability 
that the particles are found in the same state is equal to $\cos^2 \phi/2$. 
Whereas before we had a single particle interference effect, we now have a 
{\em two-particle interference}, in the sense that no measurements on 
individual particles reveal the interference phase $\phi$, while combining 
measurements on both particles makes the interference `fringe' $\cos^2 
\phi/2$ observable. 

The above argument was extended by Greenberger {\em et al.} 1990, so that 
one speaks of an ``$n$-particle interference,'' meaning a state of $n$ 
particles in which {\em no} measurements on {\em any} subset of the $n$ 
particles (containing 1, 2 or any number up to $n-1$ particles) will 
suffise to reveal an interference, but once all $n$ are measured (in the 
correct basis), and correlations established between the results, the 
interference becomes apparent. Such an $n$-particle interference is the 
state 
  \begin{equation}
\ket{n,\phi} = \left( | 000 \stackrel{n}{\cdots} \left. 0 \right>
  + \exp(i\phi )  | 111 \stackrel{n}{\cdots} \left. 1 \right>
  \right) / \sqrt{2},
  \label{nphi}  \end{equation}
where there are $n$ zeroes or ones in the ket labels, and the usual 
convention has been followed of writing product states ($\ket{0} \otimes 
\ket{0} \otimes \cdots$) by the notation $\ket{00\cdots}$. When $n=2$ the 
correlations are subject to the most simple type of Bell inequality. When 
$n=3$ we have the `GHZ' state of Greenberger, Horne and Zeilinger 1989, in 
which correlations can be found which are both non-local and which occur 
with certainty. Also, Zurek 1981 stressed that three particles are sufficient 
and necessary to establish a `preferred' basis for inter-particle 
correlations. For larger $n$, Mermin 1990 derived a Bell-type inequality 
which becomes more and more severe as $n$ grows. 

Is there a simple way of seeing that the state $\ket{n,\phi}$ is an 
$n$-particle interference? Clearly, a `which path' argument will suffice. 
If any set of less than $n$ particles is measured, the remaining unmeasured 
two-state system could in principle be measured in the $\{ \ket{0}, \ket{1} 
\}$ basis, thus indicating which of the two `paths' $|000 
\stackrel{n}{\cdots} \left. 0\right> $ or $| 111 \stackrel{n}{\cdots} 
\left. 1 \right> $ the whole system followed, which prevents any 
interference between those paths\footnote{If this description in terms of 
`following a path' is felt to be too reliant on an assumption of 
wavefunction collapse, it can always be stated more elaborately in terms of 
entanglements with external measuring devices, and the same conclusion is 
obtained.}. 

In the case that all $n$ particles are measured so as to observe a 
$\phi$-dependent result, it is instructive to examine how such interference 
can come about, it being a property of all $n$ particles, and not of any 
subset. To this end, a simple notation will be introduced. The pair of 
states $\{\ket{0}, \ket{1}\}$ will be referred to as `basis~1', and
written using standard (unbarred) labels. The states $\bet{0} \equiv 
(\ket{0} + \ket{1}) / \sqrt{2}$, and $\bet{1} \equiv (\ket{0} - \ket{1}) / 
\sqrt{2}$, will be referred to as `basis~2', and distinguished
from basis 1 by using bars over the ket symbols.
(The two bases are related by a rotation in Hilbert space through 45 
degrees). 

Consider the three-particle interference $\ket{3, \phi} = (\ket{000} + 
\exp(i\phi ) \ket{111}) / \sqrt{2}$. To observe the interference, 
measurements must be carried out in basis 2 on all the particles. 
Therefore, it is useful to write the state $\ket{3, \phi}$ in terms of 
basis states of basis 2: 
  \begin{eqnarray}
\ket{3, \phi} &\equiv& \left(1 + e^{i\phi}\right)
\left(\bet{000} + \bet{011} + \bet{101} + \bet{110} \right)/4 \nonumber \\
& & + \left(1 - e^{i\phi}\right)
\left(\bet{111} + \bet{100} + \bet{010} + \bet{001} \right)/4
  \label{3phi}  \end{eqnarray}
Measurements carried out in basis 2 will collapse the state onto one of the 
8 product states $\bet{000}$, $\bet{001} \cdots \bet{111}$.
Examining equation (\ref{3phi}), one finds 
that {\em the probability of obtaining an even number of 1's} is equal to 
$\cos^2 \phi/2$. In other words, the information on the value of $\phi$ is 
contained in the {\em parity check} of the total state in basis 2. We are 
using the word `parity' in the sense of the parity check for binary 
communication channels. 

With this insight in terms of parity, a new way of explaining the 
$n$-particle inteference arises. For, to learn the parity of a string of 
$n$ bits, it is obvious that one must know the value of all $n$ bits. No 
subset of the bits contains this information. It is important to note that 
when $\phi=0$, {\em all} the 4 possible product states of even parity 
appear, and when $\phi=\pi$, all the 4 possible product states of odd 
parity appear. If this were not the case, then sometimes a subset 
containing less than 3 bits would define the parity. For example, if we 
know from the outset that the product state $\bet{111}$ is not present in 
the final superposition, then whenever measurements of the first two bits 
both yield $1$, we know immediately that the overall parity is even, 
without measuring the third bit. 

The parity check argument is true for any $n$ (this was shown by Steane 
1996a and will also be demonstrated below). The parity check is a 
two-valued quantity, and thus can store a single bit of information. It 
may be imagined as a single bit stored symetrically among all the $n$ bits. 

\subsection{Multiple parity checks}  \label{sec:mcheck}

Once the $n$-particle interference has been understood as a parity check in 
basis 2, the concept of multiple-particle interference can be generalised. 
For, an overall parity check is the simplest example of 
{\em error detection} in a classical communication channel. More advanced 
types of error detection and correction are associated with more 
complicated types of $n$-particle interferences. To understand the details, 
we make use of theorems 1 to 3 of Steane 1996a, which are reproduced below. 
Before they are presented, a few notations will be introduced. 

First, the two-state systems which have so far been referred to as 
`particles' will hereafter be called qubits\footnote{The word `qubit'
for `quantum bit' is now a standard term for a two-state system.}.
The product states in either of bases 1 or 2 (e.g. 
$\ket{0010}$ or $\bet{01101}$) will be referred to as {\em words}, since 
each such state is identified by a unique string of bits when written in 
the relevent basis. A superposition of product states defines a set of 
words. A set of words is called a {\em code}, following standard 
nomenclature in the theory of error correcting codes. When writing 
superposition (entangled) states, the overall normalisation factor will 
often be omitted, since it is not important to the main argument, and can 
always be reintroduced easily if necessary. The theorems derived in Steane 
1996a are as follows. 

{\bf Theorem 1.}  {\em The word $\left|{000\cdots0}\right>$
consisting of all zeroes in basis $1$ is equal to 
a superposition of all $2^n$ possible words in basis $2$, with
equal coefficients.}

{\bf Theorem 2.} {\em If the $j$'th bit of each word is complemented $(0 
\leftrightarrow 1)$ in basis $1$, then all words in basis $2$ in which the 
$j$'th bit is set (is a $\bar{1}$) change sign.}

{\bf Theorem 3.} {\em When the quantum state of the system forms a linear 
code $C$ in basis $1$, in a superposition with equal coefficients, then in 
basis $2$ the words appearing in the superposition are those of the dual code 
$C^{\perp}$.} 

Theorems 1 and 2 are easy to prove, while theorem 3 requires further 
comment. A {\em linear} code $C$ is any set of $n$-bit words for which if 
the bitwise {\sc exclusive-or} (addition modulo 2) operation $\oplus$ is
carried out between 
any two words in the code, then the resulting word is also in the 
code. Such codes can be expressed in terms of an $(n \times k)$ {\em 
generator matrix} $G$, whose $k$ rows are $n$-bit words. The code consists 
of all linear combinations (by bitwise {\sc exclusive-or}) of the rows of 
$G$. This produces $2^k$ different words in the code. 
It can be shown that any linear code is also fully defined by its $(n 
\times (n-k))$ {\em parity check matrix} $H$. The code consists of all 
words $u$ for which $H_j \cdot u$ has even parity, for all rows $H_j$ of H, 
where the dot indicates the bitwise {\sc and} operation. When $H_j \cdot u$ 
has even parity, we say that ``$u$ satisfies the parity check $H_j$''. In 
other words, $H_j$ singles out a subset of the bits of $u$, and it is the 
parity of this subset which is ``checked'' when we ascertain the parity of 
$H_j \cdot u$. 

If $C$ is a linear code, then the 
dual code $C^{\perp}$ is defined to be the set of all words $v$ for which 
$v \cdot u$ has even parity for all $u \in C$. If $C^{\perp}$ is the dual 
of $C$, then $C$ is the dual of $C^{\perp}$. The only property of dual 
codes which will interest us for the moment is that {\em the generator 
matrix of a code $C$ is the parity check matrix of the dual code 
$C^{\perp}$}. (This property is used in the derivation of theorem 3, see 
Steane 1996a.) 

Using the formalism, it is possible to generalise the
concept of multiple-particle (or multiple-qubit) interference. It is 
necessary first to extend slightly the definition of the generator 
matrix. We associate with each row $G_j$ of the matrix a phase factor $\exp 
i \phi_j$, and when different rows are combined, these factors multiply: 
  \begin{equation}
G_j \oplus G_k = e^{i(\phi_j + \phi_k)} \left(
|G_j| \oplus |G_k| \right) \;\;\;\;\;\;\;\;\;\;\; (j \neq k),
  \end{equation}
where $|G_j|$ signifies the $j$'th row with phase factor set to 1. If a row 
is combined with itself, the resulting phase factor is defined to be 1, so 
that the zero word is produced: $G_j \oplus G_j = 000\cdots0$. One may 
regard the words as vectors in a discrete $n$-dimensional vector space
(Hamming space), and the phase factors as scalars. 

The generalised multiple-particle interference is defined through the 
following theorem. 

{\bf Theorem 4.} {\em If $G$ generates the state in basis 1, then the
probability that the parity check $|G_j|$ is satisfied in basis 2
varies as $\cos^2 \phi_j/2$.}

{\em Proof.} This is closely related to the proof of theorem 3. To find the 
effect of the $j$'th row of $G$, first consider the state $\ket{G'}$ 
generated by $G'$ in basis 1, where $G'$ consists of all rows of $G$ except 
the $j'th$, with all phase factors set to 1. By theorem 3, for this state, 
all the parity checks of $G'$ are satisfied in basis 2. Now complement, in 
basis 1, the qubits specified by the $j$'th row of $G$. Call the resulting 
state $\ket{G''}$. By repeated applications of theorem 2, this has the 
effect that all words change sign in basis 2 which do not satisfy the 
parity check $|G_j|$. Now form 
  \begin{equation}
\frac{\left(1 + e^{ i\phi_j}\right)}{2} (\ket{G'} + \ket{G''})
+ \frac{\left(1 - e^{i\phi_j}\right)}{2} (\ket{G'} - \ket{G''})  \equiv 
\ket{G'} + e^{i\phi_j} \ket{G''}. 
  \end{equation}
The left hand side of this equivalence shows that the probability that the 
$j$'th parity check is satisfied in basis 2 is proportional to $\cos^2 
\phi_j/2$, since if $\ket{G'}$ and $\ket{G''}$ are added (subtracted), all 
words which satisfy (respectively fail to satisfy) the parity check $|G_j|$ 
disappear in basis 2. The right hand side of the equivalence is the state 
generated by $G'$ with the row $G_j$ added to it, since the selective 
bit-complementing operation which was carried out is in fact the {\sc 
exclusive-or} operation. By applying this argument successively to all the 
rows of $G$, the theorem is proved. 

Theorem 4 is more easily understood in terms of an example, which will
now be provided. Consider the generator matrix 
  \begin{equation}
G_s =
\left( \begin{array}{ccc} e^{i \phi_1} & e^{i \phi_2} & e^{i \phi_3}
\end{array} \right) 
\left( \begin{array}{ccccccc}
0 & 0 & 0 & 1 & 1 & 1 & 1 \\
0 & 1 & 1 & 0 & 0 & 1 & 1 \\
1 & 0 & 1 & 0 & 1 & 0 & 1
\end{array} \right).
   \label{Gsphi}  \end{equation}
This equation is to be understood as a $3$-column matrix of
phase factors multiplying a $3$-row matrix of $7$-bit words.
$G_s$ generates the state 
  \begin{eqnarray}
\ket{G_s} &=&  \ket{0000000} + e^{i\phi_3}\ket{1010101} + e^{i\phi_2}
\left( \ket{0110011} + e^{i\phi_3}\ket{1100110} \right) \nonumber \\
&& + e^{i\phi_1} \left\{ \ket{0001111} + e^{i\phi_3}\ket{1011010}
+ e^{i\phi_2} \left( \ket{0111100} + e^{i\phi_3} \ket{1101001}
\right) \right\}
  \label{Gsket}  \end{eqnarray}
This code, appearing in basis 1, is well known in classical coding theory. 
It is called the simplex code, since the 8 words define the 8 vertices of a 
regular simplex in $7$-dimensional space (see for example MacWilliams \& 
Sloane 1977). In the quantum mechanical context, the code appears in 
$\ket{G_s}$ as a 7-particle entanglement containing three 4-particle 
interferences. Each phase $\phi_j$ is associated with a multiple-particle 
correlation among all those qubits which are selected by the $j$'th row of 
$G_s$. Thus, by examining the matrix $G_s$ (equation (\ref{Gsphi})), one 
sees that for this example case there are 3 correlations, each involving a 
different 4-member subset of the 7 qubits. These correlations can be 
revealed by measuring the qubits in basis 2, and calculating the relevent 
parity checks. I conjecture that such correlations satisfy Bell-type 
inequalities similar to those deduced by Mermin 1990, though a 
demonstration is beyond the scope of the present work. One may regard the 
linear codes as a generalisation to many qubits of the 2-qubit ``Bell 
basis'' $\{\ket{00} \pm \ket{11}, \ket{01} \pm \ket{10}\}$. 

This concludes the discussion of multiple particle interference {\em per 
se}. The concepts introduced make a natural introduction to the following 
sections, which will provide more information on these interferences, such 
as a method for their generation, while discussing other issues.

\section{Error Correction for Qubits}  \label{sec:ecq}

The set of $n$ qubits which we have been discussing may be considered to be 
a quantum computer (Deutsch 1985). In the course of a computation, 
entangled states involving many qubits at once are produced, and one of the 
fundamental problems of quantum computation is that such entanglements are 
highly sensitive to decoherence. `Decoherence' refers (cf Zurek 1993) to
one class of departures of the 
quantum state of the computer from the state which it ought to have. We will
not assume that decoherence is the only type of error process, however.
Rather, we seek to correct the computer (or quantum channel) in the
presence of completely general unknown departures from the state the
computer ought to have (that produced purely by evolution under
error-free computing operations)

An erroneous state of the whole computer will in general require correction 
methods operating on the whole computer at once in order to correct it. A 
method of this type was presented by Berthiame {\em et al.} 1994. Some 
types of error can be corrected through a bit-by-bit method, on the other 
hand, in which operations only on small subsets of the qubits are required. 
Shor 1995 proved that 9 qubits can be used to protect a single bit of 
quantum information against single errors, and Steane 1995 introduced a 
7-qubit scheme, and a general method for correcting many errors, while 
discussing limitations to robust encoding of a single qubit. The approach 
adopted in the latter work is generalised in this paper to enable robust 
encoding of a whole computer. Also, the method 
of how to carry out error correction without disrupting the unitary 
evolution of the computation process is given.

The philosophy pursued in this paper is to adopt methods suggested by the 
classical theory of error correction, and then to consider afterwards what 
types of error can be corrected by such methods.
It will be argued that realistic physical systems 
can be found which are subject primarily to the type of error whose 
correction we discuss. The general scenario is 
that of a computer undergoing its normal computing operations, and 
interspersed among these are error correction operations.  The qubits are 
assumed to decohere and generally change their state in an unpredictable 
manner. The word `error' will sometimes refer to a rotation of a qubit 
through $\pi/2$ radians about a given axis in Hilbert space, which mimics the 
classical `error' where a bit is complemented, but in general the word will 
refer to any departure of a qubit from the state it ought to be in. 

A general error of a qubit can be considered as made up of phase
error in basis 1 (a rotation around the axis of the Poincar\'e
or Bloch sphere) plus an amplitude error in 
basis 1 (a rotation to different latitude on the sphere), plus a 
contribution due to entanglement with external systems, which, once those 
external degrees of freedom are traced over, causes the qubit's state to 
become mixed rather than pure. We will consider first the case of phase 
error alone, then a restricted class of external entanglements, and then 
more general errors. 

\subsection{Simplest case}  \label{sec:c01}

Let us begin by considering the case that the qubits randomly dephase but
never entangle with the environment, and never flip in basis 1.
This is the simplest non-trivial case, and is practically 
interesting because it may be possible to approximate it experimentally. In 
this simple situation, the only errors are phase errors in basis 1. 

The errors are modelled by rotating the $j$'th qubit using an operator
  \begin{equation}
\left( \begin{array}{cc} e^{i \epsilon \phi_j/2} & 0 \\
0 & e^{-i \epsilon \phi_j/2}   \end{array} \right)
  \label{errmat}  \end{equation}
where the matrix has been written in basis 1. The angles $\phi_j$ are 
independent, and $\epsilon$ is a parameter used to indicate the typical 
magnitude of the errors, $0 < \epsilon \le 1$. If the single qubit state $a 
\ket{0} + b \ket{1}$ is subject to such errors, its density matrix becomes 
  \begin{eqnarray}
\rho &=& \left( \begin{array}{cc}
|a|^2              &   \alpha a b^* \\
\alpha^* a^* b     &   |b|^2 
\end{array} \right)      \label{errrho}  \\
\mbox{where } \alpha &=& e^{i \epsilon \phi}.
  \end{eqnarray}
Since $\exp (i\epsilon \phi) = 1 + O(\epsilon)$, the error in the off-diagonal 
elements is of order~$\epsilon$. 

Phase errors in basis 1 can be corrected as follows. Each qubit in the 
quantum computer is `encoded' using a set of three physical qubits, by the 
encoding method shown in figure \ref{fig:enc01}. This set is then 
`corrected' from time to time by the correction method shown in figure 
\ref{fig:cor01}. Computing operations, when required, can be carried out by 
a network equivalent to one which
first `decodes', then carries out the relevent operation, then encodes
again. The decoding operation is the inverse of the encoding one. 

To understand the error-correction scheme, one notes that it is based on 
the simplest classical error correction code, the $n=3$ repetition code, 
operating in basis 2. This is because phase errors in basis 1 cause 
amplitude errors in basis 2, so we employ a scheme which corrects amplitude 
errors in basis 2. A general single-qubit state $a \ket{0} + b \ket{1}$ is 
encoded by two controlled not ({\sc cnot}) operations in
basis~2,\footnote{ {\sc cnot} in basis 1 is $(
\left| 00\right> \left< 00\right| + 
\left| 01\right> \left< 01\right| + 
\left| 10\right> \left< 11\right| + 
\left| 11\right> \left< 10\right|)$ where the first qubit is the control, 
the second is the target. {\sc cnot} in basis 2 (as here) is the operator 
having the same form, but with $0$ and $1$ replaced by $\bar{0}$ and
$\bar{1}$.} 
acting on an initial state $(a \ket{0} + b \ket{1}) \otimes \bet{00}$ (see 
figure \ref{fig:enc01}). The state thus encoded using three qubits is 
  \begin{eqnarray}
&& a \left(\ket{000} + \ket{011} + \ket{101} + \ket{110} \right)
+ b \left(\ket{111} + \ket{100} + \ket{010} + \ket{001} \right) \nonumber \\
&=& (a + b) \bet{000} + (a-b) \bet{111}      \label{encstate}
  \end{eqnarray}
Therefore the only `legal' states are $\bet{000}$ and $\bet{111}$ or linear 
combinations of these.

The random phase errors in basis 1 cause departures from the subspace 
spanned by $\bet{000}$ and $\bet{111}$. As long as the errors are small, 
the component which was $\bet{000}$ is likely to remain in the region of 
Hilbert space spanned by $\{\bet{000}, \bet{001}, \bet{010}, \bet{100}\}$ 
while the component which was $\bet{111}$ is likely to remain in the region 
spanned by $\{\bet{111}, \bet{110}, \bet{101}, \bet{011}\}$. As long as 
only such `single errors' occur, they can be corrected. 

The error corrector shown in figure \ref{fig:cor01} works as follows. 
First, two {\sc cnot} operations carry out parity checks. The checks 
required are those given by the parity check matrix for the $\{000,111\}$ 
repetition code: 
  \begin{equation}
H_{\rm rep} = \left( \begin{array}{ccc} 1&1&0 \\ 1&0&1 \end{array} \right)
  \end{equation}
After these checks, the `control' qubit contains the state to be corrected, 
and the other two `target' qubits (hereafter called parity qubits) contain 
the error syndrome. The two parity qubits containing the syndrome are now 
measured. The syndrome indicates which qubit is to be complemented. That 
is, if the measurements give $11$ then the {\sc not} operation is carried 
out on the control qubit.\footnote{ {\sc not} in basis 1 is the operator 
$\ket{0}\left<1\right| + \ket{1}\left<0\right|$. {\sc not} in basis 2 (as 
here) is $\bet{0}\left< \right.\overline{1} \left.\right| +
\bet{1}\left< \right.\overline{0}\left.\right|$.} 
Whatever the result of the measurements, the parity qubits are reset to 
$\bet{00}$. After this, the three qubits are in the decoded state. The 
final part of the error corrector reencodes the state. 

The effect of the above transformations can easily be calculated. Once the 
encoding has been carried out, yielding the state given by equation 
(\ref{encstate}), all three bits are subjected to errors given by the 
operator (\ref{errmat}) with independent unknown $\phi_0, \phi_1,\phi_2$. 
The resulting erroneous state is to be corrected. The two {\sc cnot} 
operations are applied, and measurements are modelled by projection 
operators. This yields 4 different density matrices for the 4 different 
measurement outcomes. The {\sc not} operation is carried out on the 
relevent qubit or bits as indicated by the syndrome associated with each 
density matrix. The resulting four density matrices are added with the 
weights given by the probabilities of obtaining them. Now we are at the 
stage just before the final reencoding. If instead of reencoding, we simply 
extract the density matrix of the control qubit, the result is equation
(\ref{errrho}) with
  \begin{eqnarray}
\alpha &=& \frac{1}{2}\left\{
\cos (\epsilon \phi_0) + \cos (\epsilon \phi_1) + \cos( \epsilon \phi_2)
\right. \nonumber \\
&& \left.
- \cos (\epsilon \phi_0) \cos (\epsilon \phi_1) \cos( \epsilon \phi_2)
- i \sin (\epsilon \phi_0) \sin (\epsilon \phi_1) \sin( \epsilon \phi_2)
\right\}.
  \end{eqnarray}
When only one of the three angles $\phi_j$ is non-zero (that is, one qubit 
is erroneous), the state is restored exactly, and when all three are 
non-zero, the error term is of order $\epsilon^3$ instead of order 
$\epsilon$, as a Taylor expansion of the trigonometric functions will show 
($\alpha = 1 + O(\epsilon^3)$). The corresponding properties of classical 
single-error correction are that single errors are corrected exactly, and 
error probabilities of order $p$ become of order $p^2$ after correction. 
Since in this case the qubit error term goes directly to $O(\epsilon^3)$ 
rather than $O(\epsilon^2)$, the correction is efficient. In the next 
section a case which mimics the classical behaviour more closely will be 
discussed. 

It has been assumed throughout that the process of encoding and
correcting does not itself introduce more errors than it corrects. 

The discussion so far only demonstrates a modest correction ability. 
However, the concepts can be generalised, enabling the
limitations of the method to be derived. We turn
to this in later sections. The main result so far 
is to show that unitary evolution of a qubit can be preserved, while 
information about error processes is nevertheless gathered and used to 
correct the qubit. Next it will be shown that the general methods discussed 
in this paper are not limited to the correction of unitary errors, but can 
enable the quantum computer to recover from relaxation caused by erroneous 
coupling to its environment. 

\subsection{Simplest Purity Amplification}  \label{sec:spa}

The single error correction in basis 2 discussed in the previous section 
with regard to unitary phase errors in basis 1 is also sufficient to 
correct a restricted class of relaxation errors (ie errors caused
by coupling to external systems). 
The restricted class is relaxation which does not cause amplitude errors in 
basis 1. One can model such relaxation either as a decay in the 
off-diagonal density matrix elements of each qubit in basis 1, or as an 
entanglement with the environment introduced by operators of the type 
  \begin{equation}
W_j = \begin{array}{c}
\ket{0}_j \otimes \ket{ \psi_{1j} } \\
\ket{0}_j \otimes \ket{ \psi_{2j} } \\
\ket{1}_j \otimes \ket{ \psi_{1j} } \\
\ket{1}_j \otimes \ket{ \psi_{2j} }
\end{array}
\; \;
\left( \begin{array}{cccc}
1 & 0 & 0 & 0 \\
0 & 1 & 0 & 0 \\
0 & 0 & 1 - \epsilon_j                & \sqrt{ 2\epsilon_j-\epsilon_j^2 } \\
0 & 0 & -\sqrt{ 2\epsilon_j-\epsilon_j^2 } & 1 - \epsilon_j
\end{array} \right).
  \label{entangle}  \end{equation}
Here, $\ket{\psi_{1j}}$ and $\ket{\psi_{2j}}$ are orthogonal states of the 
environment, and the product states on the left of the operator matrix 
indicate the basis in which the matrix is written. The (real) parameter 
$\epsilon_j$, bounded by $0 < \epsilon_j \leq 1$, indicates the strength of 
the entanglement with the environment. The entanglement can be imagined as 
an imperfect ($\epsilon < 1$) or perfect ($\epsilon = 1$) measurement of 
the qubit in basis 1. Such entanglements, when perfect, have the effect of 
making the ``interference phase'' between the two parts $\ket{0}$ and 
$\ket{1}$ of qubit's state unobservable, as a straightforward analysis
will show. No amplitude error is
introduced in basis 1, which is the clue that
the `simplest possible' error-correction procedure 
introduced in the previous section will be sufficient to correct errors 
having this form. 

The effect of the entanglement $W_j$ on a single qubit is calculated
by operating $W_j$ on the joint qubit--environment state $(a \ket{0}
+ b \ket{1}) \otimes \ket{ \psi_1 }$, and then obtaining the reduced
density matrix of the qubit by tracing over the environment variables.
The result is a density matrix as in equation (\ref{errrho}), with
  \begin{equation}
\alpha = 1-\epsilon.
  \label{enterr}  \end{equation}
This is clearly a mixed state when $\epsilon > 0$, and the error
term is of $O(\epsilon)$.

When we examine the density matrix in basis 2, this error appears partly as 
an amplitude error, and it can be corrected by the encoding and correcting 
procedure described in the previous section (figures \ref{fig:enc01} and 
\ref{fig:cor01}). To calculate the effects, first the general single-qubit 
state $(a \ket{0} + b \ket{1})$ is encoded using three qubits, then each of 
the three undergoes entanglement with the environment, described by three 
operators $W_0, W_1, W_2$ defined by equation \ref{entangle}.
The overall state 
then involves 8 different environment states, associated with 8 different 
3-qubit states. The error correction procedure is carried out next. In the 
calculation, it appears as a set of eight independent corrections on each 
of the eight 3-qubit states. The final `corrected' 3-qubit density matrix 
is then taken to be the weighted sum of the eight 3-qubit density matrices 
associated with different states of the environment. The density matrix of 
the control qubit is extracted, yielding the form (\ref{errrho}) with
  \begin{equation}
\alpha = 1 - \frac{1}{2} \left(
\epsilon_0 \epsilon_1 + \epsilon_0 \epsilon_2 + \epsilon_1 \epsilon_2
\right) + \frac{1}{2} \epsilon_0 \epsilon_1 \epsilon_2 
  \label{entcor}  \end{equation}
This result shows that when only a single qubit decoheres (ie only 
one of the entanglement terms $\epsilon_j$ is non-zero), the state is 
corrected exactly ($\alpha=1$), and when all three undergo errors, the
error term is of 
$O(\epsilon^2)$ instead of $O(\epsilon)$. The corresponding properties of 
classical single-error correction are that single errors are corrected 
exactly, and error probabilities of order $p$ become of order $p^2$ after 
correction. 

The fact that the corrected density matrix is 
nearer to `pure state' conditions than the original density matrix
(when $\epsilon = \epsilon_j < 1$), is an example of a 
general phenomenon called `quantum privacy amplification' in the context of 
a quantum communication channel (Bennett {\em et al.} 1995; Deutsch {\em et 
al} 1996), and which will be referred to here as `purity amplification'. 
The ability to implement purity amplification is an important part of the 
general problem of error correction in quantum communication channels and 
computers. This section has shown that the approach to error correction 
adopted in this paper is capable of handling purity amplification. Indeed, 
the `quantum privacy amplification' protocol described by Deutsch {\em et 
al.} 1996 can be understood as an implementation of single-error {\em 
detection} in basis 1 and basis 2 simultaneously, by means of a single 
parity check in each basis. The fact that it is a detection rather than 
correction scheme explains why non-useful pairs of bits have to be thrown 
away. 

\subsection{General single error correction} \label{sec:c11}

Suppose now the type of error is completely general---there
is an arbitrary change in the state of a qubit, including possible relaxation.
To correct this, the method is to implement single error
correction in both bases 
simultaneously. For this, the encoding used in the previous section is not 
sufficient, since there single errors in basis 1 could not be corrected. To 
understand the encoding requirements, the concept of {\em minimum 
distance}, introduced by Hamming 1950, is employed.
The Hamming distance between two words is 
equal to the number of bits which must be complemented in order to convert 
one word into the other. The minimum distance $d$ of a code is the minimum 
Hamming distance between any two words in the code. A code of minimum 
distance $d > 2 x$ is necessary if $x$ errors are to be corrected, since
only if fewer than $d/2$ errors occur can the codeword which gave rise to the 
erroneous word can be identified unambiguously as the only codeword within 
distance $d/2$ of the erroneous word. In what follows, the standard 
notation $[n,k,d]$ will be employed to refer to a linear code using $n$ 
bits, having $2^k$ codewords and minimum distance $d$. 

To correct for a general single error, we require an encoding allowing 
minimum distance 3 in both basis 1 and basis 2. A method to do this was 
presented by Steane 1996a, as follows. We seek a code $C$ having the 
following properties: its dual code $C^{\perp}$ has minimum distance 3, and 
it is itself a subcode of a of code $C^{+}$ of minimum distance 3. The 
reasoning behind this is best demonstrated by means of an example. 

The $[7,3,4]$ simplex code presented in section \ref{sec:mcheck} has the 
properties required. Its dual code is the $[7,4,3]$ Hamming code which has 
minimum distance 3, and it is a subcode of the $[7,4,3]$ punctured 
Reed-Muller code, also of minimum distance 3. $n=7$ is the smallest number 
of bits for which a code can be found with these properties. The encoding 
and correcting procedure is shown in figures \ref{fig:enc11} and 
\ref{fig:cor11}, and explained as follows. 

The encoding method is based on the generator matrix of the $[7,3,4]$ 
simplex code, given by equation (\ref{Gsphi}) with $\phi_1, \phi_2,
\phi_3 = 0$.
The simplex code thus generated will be called $\ket{C}$. It is the state 
$\ket{G_s}$ shown in equation (\ref{Gsket}) with all phase angles set to 
zero. The essential idea is that a qubit state $\ket{0}$ is encoded as 
$\ket{C}$, while a qubit state $\ket{1}$ is encoded as $\ket{\neg C}$, 
which is $\ket{C}$ with the {\sc not} operation carried out on all the 
qubits, ie the coset $\ket{C \oplus 1111111}$.
It is easy to deduce that when the qubit $Q$ to be encoded is in 
the state $\ket{0}$, the encoder shown in figure \ref{fig:enc11} places the 
7 qubits in the state $\ket{C}$. To see that $\ket{1}$ becomes encoded as 
$\ket{\neg C}$, consider the parity check matrix of $\ket{C}$: 
  \begin{equation}
H_C = \left(  \begin{array}{ccccccc}
1 & 1 & 0 & 1 & 0 & 0 & 1 \\
0 & 1 & 0 & 1 & 0 & 1 & 0 \\
1 & 0 & 0 & 1 & 1 & 0 & 0 \\
1 & 1 & 1 & 0 & 0 & 0 & 0
\end{array} \right)
  \end{equation}
Now, $\ket{\neg C}$ fails all those parity checks for which there is an odd 
number of $1$'s in the relevent row of $H_s$, and passes the others. Hence, 
the operations which generate $\ket{C}$ when starting from $\ket{0000000}$, 
will generate $\ket{\neg C}$ when starting from $\ket{0010110}$, since the 
complemented qubits ensure that the final state will pass and fail the 
checks in $H_C$ in the way appropriate for $\ket{\neg C}$. This initial 
complementing of qubits is the job of the first two {\sc cnot} operations 
in the encoder. The encoder therefore encodes a general single qubit state 
$(a\ket{0} + b\ket{1})$ as $(a\ket{C} + b \ket{\neg C})$. 

Now, $\ket{C}$ and $\ket{\neg C}$ are subcodes (strictly, cosets)
of the $[7,4,3]$
punctured Reed-Muller code $C^+$ whose parity check matrix is
  \begin{equation}
H_{C+} = \left( \begin{array}{ccccccc}
0 & 1 & 1 & 1 & 1 & 0 & 0 \\
1 & 0 & 1 & 1 & 0 & 1 & 0 \\
1 & 1 & 0 & 1 & 0 & 0 & 1
\end{array} \right)
  \end{equation}
All legal encoded states satisfy this parity check matrix, and this is the 
basis of the corrector shown in figure \ref{fig:cor11}. The operation of 
the corrector is similar to that of the corrector presented in the previous 
section (figure \ref{fig:cor01}). Multiple {\sc cnot} operations are used 
to carry out parity checks, the parity qubits are measured, and the state 
corrected by means of {\sc not} operations on qubits identified by the 
measured syndrome. Finally, the state is reencoded. 

So far, error correction has been carried out in basis 1. However, the 
coding was carefully selected in such a way that only words in the 
$[7,4,3]$ Hamming code $C^{\perp}$ should appear in basis 2. Therefore, 
error correction can also be carried out in basis 2. The corrector is based 
on the parity check matrix of the $[7,4,3]$ Hamming code, which is equal to 
the generator matrix of its dual, the matrix $G_C$ given in equation 
(\ref{Gsphi}). Now, it so happens that the $[7,4,3]$ Hamming code is
the same as the $[7,4,3]$ punctured Reed-Muller
code (this is not always true for higher 
order codes), as can be seen by the fact that $G_C$ and $H_{C+}$ are equal 
(one can be converted to the other by linearly combining rows). Therefore, 
the corrector in basis 2 is once again given by figure \ref{fig:cor11}, 
only now all the operations are carried out in basis 2. 

The correction scheme described will tend to keep the encoded state 
confined to the region of Hilbert space spanned by the two state vectors 
$\{\ket{C}, \ket{\neg C}\}$. This is a two-dimensional subspace within the 
128-dimensional total Hilbert space. The subspace is also spanned by the 
state vectors $\{\bet{H,e}, \bet{H,o}\}$ defined by the even and odd parity 
subcodes of the $[7,4,3]$ Hamming code in basis 2, since theorem 4 implies 
that $\ket{C} + \ket{\neg C} \equiv \bet{H,e}$ and $\ket{C} - \ket{\neg C} 
\equiv \bet{H,o}$, up to a normalisation factor. If an arbitrary 
single-qubit state is encoded as $(a\ket{C} + b \ket{\neg C})$, and then 
any one (but only one) of the 7 qubits is allowed to change state and 
entangle with the environment in an arbitrary manner, the error corrector described in this 
section will return the 7 qubits exactly to the error-free state $(a\ket{C} 
+ b \ket{\neg C})$. This will be proved below as part of the more general
theorem 6.
If more than one qubit is allowed to undergo errors, 
then error terms which would be of order $\epsilon$ in the density matrix 
of an uncorrected qubit become of order $\epsilon^2$ or higher when 
encoding and correction is employed. 

\subsection{Multiple correction of multiple qubits} \label{sec:mec}

The previous sections have introduced almost sufficient insights to enable 
the general problem of multiple error correction of many qubits to be 
addressed. The final ingredient is theorem 5 below. Before it is presented, 
we remark that just as in classical information theory, it is necessary to 
distinguish between the amount of information $k$ and the number of bits 
$n$ used in a $[n,k,d]$ code, it will be necessary here to distinguish 
between the number $K$ of independent quantum bits of information we wish 
to keep free of errors, and the number $n$ of qubits used to do this. Thus, 
in section \ref{sec:c01} a single qubit was encoded, $K=1$, by means of 
$n=3$ encoding qubits, and in section \ref{sec:c11} a single qubit $K=1$ 
was encoded by means of $n=7$ encoding qubits. 

{\bf Theorem 5.} {\em To encode $K$ qubits with minimum distance $d_1$
in one basis, and minimum distance $d_2$ in the other, it is sufficient
to find a linear code $C^{+K}$ of minimum distance $d_1$, whose $K$'th
order subcode $C$ is the dual of a distance $d_2$ code.}\\
{\bf Corollary:} {\em Finding such a code is sufficient not only to 
demonstrate that the encoding is possible, but also to make self-evident 
the physical procedures for encoding and correction.} 

A $K$'th order subcode of $C^{+K}$ is a code obtained by adding $K$ rows 
to the parity check matrix of $C^{+K}$. 

{\em Proof:} The general insight is that whereas in classical theory, 
information is encoded using different {\em words} of a code, in the 
quantum mechanical case, information is encoded using different {\em 
cosets} of a code. Thus in section
\ref{sec:c11}, cosets of the $[7,4,3]$ punctured Reed-Muller 
code, were used, and in section \ref{sec:c01}, the even parity and
odd parity codes in basis 1 were 
cosets of the $[n,n,1]$ code of all possible words. 

To encode $K$ qubits, we require $2^K$ cosets. If these are all 
non-overlapping cosets of a distance $d_1$ code, then clearly they are 
all separated from one another by at least $d_1$. That is, all words in one 
coset are at least $d_1$ from all words in another coset. We can ensure 
the cosets do not overlap by defining them as follows. $K$ new rows are 
added to the parity check matrix of $C^{+K}$. The new rows are linearly 
independent of each other and of all the other rows. (If this is not 
possible then $n$ must be increased and the argument restarted). Each of 
the $K$ new parity checks can either be satisfied or not satisfied. This 
allows $2^K$ different possibilities, each of which produces a coset 
which has no words in common with any of the other cosets. Hence the 
cosets are non-overlapping. 

Suppose the first coset $C$ is a code having a dual $C^{\perp}$. To
obtain one of 
the other $2^K - 1$ cosets from $C$, it is sufficient to complement 
in basis 1 whichever parity qubits implement a parity check which $C$
satisfies but the new coset does not. The
effect in the other basis is to change the 
sign of some of the words (by theorem 2).
(Equivalently, it is sufficient to
change the sign of the relevent rows of the parity check matrix in
basis 1, which is the generator matrix in basis 2, by theorem 4).
Therefore, each coset in basis 
1 produces the words of $C^{\perp}$ in basis 2, with signs depending on the 
coset. Hence, for any superposition of the cosets in basis 1, all words 
appearing in basis 2 are in the code $C^{\perp}$. Therefore, if we require 
minimum distance $d_2$ in basis 2, it is sufficient that $C^{\perp}$ should 
be a distance $d_2$ code, and the theorem is proved. 

The coding method of theorem 5 uses a $2^K$-dimensional subspace to store 
the quantum information, within a total Hilbert space of dimension $2^n$. 
The subspace is spanned by the $2^K$ cosets of $C^{+K}$ in basis 1, and 
by $2^K$ cosets of $C^{\perp}$ in basis 2. The encoding and correcting 
operations are deduced directly from the parity check matrices of the 
relevant linear codes, in the manner illustrated by figures \ref{fig:enc01} 
to \ref{fig:cor11}. An alternative approach to error correction is illustrated
by figure \ref{fig:anc}. 
To implement correction in basis 1 (basis 2), a set
of $n-k_1$ (respectively $n-k_2$)
ancillary qubits is introduced, and
the error syndrome is stored into this ancilla by means of multiple
{\sc cnot} operations. The operations required are exactly those specified
by the parity check matrix of $C^{+K}$ (respectively $C^{\perp}$),
which proves the corollary to theorem 5.
The ancilla is measured (in the relevent basis), and
the result used to calculate which qubits in the quantum
computer are to undergo a {\sc not} operation. 

\subsection{Error correction in two bases is sufficient}

This section is dedicated to the proof of the following theorem.

{\bf Theorem 6.} {\em Error correction in basis 1 followed by error correction
in basis 2 is sufficient to restore the quantum computer after arbitrary
errors of a small enough subset of its qubits. Specifically,
if $x$ qubits undergo errors, then correction is successful
if at least $x$ errors can be corrected in both basis 1 and basis 2.}

This theorem shows that the correction methods described in this paper are
not limited to the correction of simple `qubit complementing' errors, but 
can handle any error process, as long as it only affects a subset of the
$n$ qubits in the computer. To keep a clear distinction, the word {\em flip}
is reserved in this section to refer to error processes of the form 
either $\ket{0} \leftrightarrow \ket{1}$ (`a flip in basis 1') or $\bet{0} 
\leftrightarrow \bet{1}$ (`a flip in basis 2'). Completely general erroneous 
changes in the state of a qubit, including entanglement with the environment, 
will be referred to as {\em defection}\footnote{Random unitary changes
(rotations) of a qubit are caused by defects in the
quantum computer; to entangle randomly with the environment is to form a treacherous alliance with an enemy of successful quantum computation.}. 

The following notations will be used.
  \begin{eqnarray*}
\ket{Ci} &=& \mbox{the $i$'th coset of } \ket{C^{+K}}. \\
\ket{Ci_j} &=& \mbox{the $j$'th coset of }\ket{Ci}. \\
\ket{Ci_j / \,^1\!S_k} &=& \ket{Ci_j} \mbox{ subject to flips in basis 1
whose error syndrome is} S_k \\
\ket{Ci / \,^2\!S_l} &=& \ket{Ci} \mbox{ subject to flips in basis 2
whose error syndrome is } S_l \\
\ket{e_n} &=& \mbox{ a state of the environment.}
  \end{eqnarray*}
As an example of the above, consider the state
$\ket{Ci} = \ket{0000} + \ket{0011} + \ket{1100} + \ket{1111}$.
One pair of possible cosets is $\ket{Ci_0} = \ket{0000} + \ket{0011}$
and $\ket{Ci_1} = \ket{1100} + \ket{1111}$. Suppose the error
syndrome for the case of no errors is $S_0$, then $\ket{Ci_j / \,^1\!S_0}
= \ket{Ci_j}$. If a single error in the last bit produces the
syndrome $S_1$, then
$\ket{Ci_0 / \,^1\!S_1} = \ket{0001} + \ket{0010}$;
$\ket{Ci_1 / \,^1\!S_1} = \ket{1101} + \ket{1110}$;
$\ket{Ci / \,^2\!S_1} = \ket{0000} - \ket{0011} + \ket{1100} - \ket{1111}$.

In what follows, we will require the following result:
  \begin{equation}
\ket{Ci_j} = \sum_{l=0}^{2^x-1} \ket{Ci / \,^2\!S_l}
(-1)^{{\rm wt}(j \cdot l)}
\label{Cij}  \end{equation}
where the notation ${\rm wt}(j \cdot l)$ means the Hamming weight\footnote{The
Hamming weight of a bit string $x$ is the number of $1$'s in $x$.}
of $j \cdot l$, and the dot indicates the bitwise {\sc and} operation
carried out between $j$ and $l$. The result holds for a particular
type of coset $Ci_j$, which will be identified shortly. 

It is obvious that a code can be written
as the sum of its cosets: $\ket{Ci} = \sum_j \ket{Ci_j}$.
The content of (\ref{Cij}) is the inverse result, that a coset $\ket{Ci_j}$ 
can be written as a sum of (erroneous) codes. The flips in basis 2 cause sign 
changes amongst the basis 1 words of $\ket{Ci}$ in such a way that when the
sum in 
equation (\ref{Cij}) is carried out, all words in $\ket{Ci}$ which do not 
belong to $\ket{Ci_j}$ cancel, so the result is $\ket{Ci_j}$. 

{\em Proof of equation (\ref{Cij})}.
If $Ci_j$ is an $x$'th order coset of $Ci$, then
the parity check matrix for $Ci_j$ consists of the
parity check matrix of $Ci$, plus $x$ extra rows. We consider the case
that each of these extra rows contains all zeroes apart from a single
$1$. For example, for $x=3$, $n=10$, the $8$ cosets might be counted
by the values of the fourth, sixth and tenth qubits in basis 1, in which case
the parity check matrix in basis 1 is
  \begin{equation}
\ket{Ci_j} \leftrightarrow \left( \begin{array}{c}
H_i \\
\hline
(-1)^{{\rm wt}(j \cdot 001)} \, 0000000001 \\
(-1)^{{\rm wt}(j \cdot 010)} \, 0000010000 \\
(-1)^{{\rm wt}(j \cdot 100)} \, 0001000000
\end{array}  \right)
\label{H+}  \end{equation}
where $H_i$ is the parity check matrix of $\ket{Ci}$.
The $j$'th coset passes or fails these extra parity checks according
as the bits of the binary value of $j$ are zero or one. This pass$/$fail
property is indicated by the sign (power of $-1$) in
front of each row of the
matrix.\footnote{This sign is an example of the phase factor introduced in
section \ref{sec:mcheck} to generalise the use of such matrices in the quantum 
mechanical as opposed to classical context (cf Theorem 4).}
Thus, $Ci_0$ is the set of words of $Ci$ for which the 4'th, 6'th and 10'th 
bits are zero, $Ci_5$ is the set of words of $Ci$ for which the 4'th and 10'th 
bits are one, and the 6'th is zero, since decimal 5 is binary 101, and so on. 

In general, the type of coset for which equation (\ref{Cij}) holds is one 
consisting of all words in $\ket{Ci}$ for which a chosen set of $x$ qubits 
has the value $j$ in basis~$1$.

Since the matrix in equation (\ref{H+}) is the parity check matrix of 
$\ket{Ci_j}$ in basis 1, it is the generator matrix in basis 2
of the same quantum state (theorem 4).
But, such a generator matrix will generate the code $Ci$ plus
$2^x-1$ erroneous copies of $Ci$, where a given copy will have flips of just
the bits selected by those extra rows of the generator matrix which were used 
to generate that copy. Hence, equation (\ref{Cij}) is proved.

We now pass on to the question of general errors and their correction.
A general defection of a single qubit can be written
  \begin{equation}
\left. \begin{array}{ccc}
\ket{0} \ket{e_0} &\rightarrow& \ket{0}\ket{e_1} + \ket{1}\ket{e_2} \\
\ket{1} \ket{e_0} &\rightarrow& \ket{0}\ket{e_3} + \ket{1}\ket{e_4}
\end{array} \right\} 
  \end{equation}
where no assumptions are made about the environment
states $\ket{e_i}$---they may
or may not be orthogonal, and they may include arbitrary (complex)
coefficients (they are not normalised). A general defection of
$x$ qubits is
  \begin{equation}
\ket{j}\ket{e_0} \rightarrow \sum_{k=0}^{2^x-1} \ket{j / \,^1\!S_k} \ket{e_{jk}}
\label{decoh}  \end{equation}
where $\ket{j}$ is any one of the $2^x$ possible $x$-qubit words.

Now, suppose that in some state $\ket{Ci}$, a subset of the qubits defect.
The subset contains $x$ qubits positioned anywhere among the $n$ qubits
of the total system. A state $\ket{Ci}$ (ie before defection) can be
written
$\ket{Ci} = \sum_{j=0}^{2^x-1} \ket{Ci_j}$
where the $j$'th coset consists of all words in $\ket{Ci}$ for which
the subset of $x$ bits has the value $j$ in basis 1. A general 
defection among the $x$ defecting qubits is the process indicated by 
equation (\ref{decoh}), with the $n-x$ unchanged qubits acting as spectators, 
and with $j$ indicating the initial values of the decohering qubits. Therefore, 
the effect of defection on $\ket{Ci_j}$ is 
  \begin{equation}
\ket{Ci_j} \ket{e_0} \rightarrow
\sum_{k=0}^{2^x-1} \ket{Ci_j / \,^1\!S_k} \ket{e_{jk}}.
\label{subdecoh}  \end{equation}
Note that the state of the environment after defection is independent
of $i$ in this equation. This is because we selected the cosets $\ket{Ci_j}$ 
in such a way as to bring exactly this property about. 
The environment does not `care' about the state of the spectator
qubits, so its final state is not sensitive to which code $\ket{Ci}$
gave rise to the coset $\ket{Ci_j}$.

We now have enough results to prove Theorem 6.

{\em The proof:} Using the encoding method of Theorem 5, a general
state of a computer before defection can be written
  \begin{equation}
\ket{QC} = \sum_{i=0}^{2^K-1} c_i \ket{Ci}
  \end{equation}
Expanding each state $\ket{Ci}$ as a set of cosets, this is
  \begin{equation}
\ket{QC} = \sum_{i=0}^{2^K-1} c_i \sum_{j=0}^{2^x-1} \ket{Ci_j}
  \end{equation}
where we choose the set of cosets identified by the $2^x$ possible values in 
basis 1 of the $x$ qubits which now defect. Using equation (\ref{subdecoh}),
the effect of defection is
  \begin{equation}
\ket{QC}\ket{e_0} \rightarrow
\sum_{i=0}^{2^K-1} c_i \sum_{j=0}^{2^x-1} 
\sum_{k=0}^{2^x-1} \ket{Ci_j / \,^1\!S_k} \ket{e_{jk}}.
  \end{equation}
Now apply error correction in basis 1. As long as
$x < d_1 / 2$, this has the effect that
  \begin{equation}
\ket{Ci_j / \,^1\!S_k} \ket{^1m_0} \rightarrow \ket{Ci_j} \ket{^1m_k}
  \end{equation}
where $\ket{^1m_k}$ indicates a state of the measuring apparatus
used for correction (cf figure \ref{fig:anc}).
Therefore the total state of the quantum
computer, environment and measuring apparatus becomes
  \begin{eqnarray}
&&
\sum_{i=0}^{2^K-1} c_i \sum_{j=0}^{2^x-1} 
\sum_{k=0}^{2^x-1} \ket{Ci_j} \ket{^1m_k} \ket{e_{jk}} \\
&=& 
\sum_{i=0}^{2^K-1} c_i \sum_{j=0}^{2^x-1} 
 \ket{Ci_j}
\left( \sum_{k=0}^{2^x-1} \ket{^1m_k, e_{jk}} \right).
\label{halfway}  \end{eqnarray}
Note that the error correction has corrected all $2^x$ cosets
$\ket{Ci_j}$ in parallel. The correction is not yet complete because
each coset is entangled with a different state of the environment.

Using equation (\ref{Cij}), the total state given by (\ref{halfway})
can be written
  \begin{equation}
\sum_{i=0}^{2^K-1} c_i \sum_{j=0}^{2^x-1} 
\sum_{l=0}^{2^x-1} \ket{Ci / \,^2\!S_l} (-1)^{{\rm wt}(j \cdot l)}
\left( \sum_{k=0}^{2^x-1} \ket{^1m_k, e_{jk}} \right).
  \end{equation}
Now apply error correction in basis 2. As long as $x < d_2 / 2$,
this has the effect that
  \begin{equation}
\ket{Ci / \,^2\!S_l} \ket{^2m_0} \rightarrow \ket{Ci} \ket{^2m_l}
  \end{equation}
where $\ket{^2m_l}$ indicates a state of the measuring apparatus
using for correction in basis 2. Therefore the total state of the
quantum computer, environment and both measuring apparati becomes
  \begin{eqnarray}
&& \sum_{i=0}^{2^K-1} c_i \sum_{j=0}^{2^x-1} 
\sum_{l=0}^{2^x-1} \ket{Ci} \ket{^2m_l} (-1)^{{\rm wt}(j \cdot l)}
\left( \sum_{k=0}^{2^x-1} \ket{^1m_k, e_{jk}} \right) \\
&=& 
\left( \sum_{i=0}^{2^K-1} c_i \ket{Ci} \right)
\sum_{j=0}^{2^x-1} 
\sum_{l=0}^{2^x-1} \ket{^2m_l} (-1)^{{\rm wt}(j \cdot l)}
\left( \sum_{k=0}^{2^x-1} \ket{^1m_k, e_{jk}} \right) \\
&=& \ket{QC} \otimes \ket{^2m, ^1m, e}.
  \end{eqnarray}
Hence, the quantum computer becomes completely disentangled from
its environment, and is returned to its initial state. Therefore,
error correction in basis 1 and basis 2 is sufficient to
restore the quantum computer after arbitrary errors of
$x < d_1/2, \, d_2/2$ qubits, and theorem 6 is proved.

\section{Error rate limitations}  \label{sec:errlim}

We now turn to the question of whether errors can be supressed sufficiently 
to enable useful computations to be carried out on a quantum computer. This 
may also be regarded as a problem of communication over a noisy quantum 
channel. The method is to establish the limitations implicit in the coding 
method described by theorems 5 and 6. 

The fundamental problem of the theory of classical error correcting codes 
is to find codes of length $n$ (ie $n$ is the length of the words) and 
minimum distance $d$ which contain the maximum possible number of 
codewords. Let this maximum possible number of codewords be $A(n,d)$. 
Although $A(n,d)$ is not known in general, a number of upper and lower 
bounds have been established. In what follows, we will make use of two 
simple bounds. The first is the Hamming or sphere-packing bound introduced 
by Hamming 1950. In the limit of large $n$, it takes the form
  \begin{equation}
\frac{\log_2(A(n,d))}{n} \leq \left( 1 - H\left(\frac{d}{2n}\right) \right)
\left(1 - \zeta \right)
\label{Hammn}  \end{equation}
where $\zeta \rightarrow 0$ as $n \rightarrow \infty$, and $H(x)$ is the 
entropy function 
  \begin{equation}
H(x) \equiv x \log_2 \frac{1}{x} + (1-x) \log_2 \frac {1}{1-x}.
  \end{equation}
There are no codes of length $n$ and 
distance $d$ which have more words than this upper limit, and usually the 
upper bound itself cannot be achieved. This is the ``bad news''. The good 
news is that useful codes do exist. 
The Gilbert-Varshamov bound (Gilbert 1952; Varshamov 1957; see also 
MacWilliams \& Sloane 1977) is a sufficient but not necessary condition 
for the existence of a $[n,k,d]$ code. In the limit of large $n$, it takes
the form
  \begin{equation}
\frac{k}{n} \ge \left( 1 - H\left( \frac{d}{n} \right) \right)
\left(1 - \zeta \right)
\label{GVn}  \end{equation}
where $\zeta \rightarrow 0$ as $n \rightarrow \infty$. It can be shown 
(MacWilliams \& Sloane 1977) that there exists an infinite sequence of 
$[n,k,d]$ linear codes satisfying inequality (\ref{GVn}) with $d/n \ge 
\delta$ if $0 \le \delta < 1/2$. 

Theorem 5 states that to encode $K$ qubits with minimum distances $d_1$ and 
$d_2$ in bases 1 and 2, we require codes $C^{+K}$, $C$, $C^{\perp}$ related 
as follows: 
  \begin{equation}
[n,x+K,d_1] \stackrel{\rm subcode}{\longrightarrow} 
[n,x,y] \stackrel{\rm dual}{\longleftrightarrow} [n,n-x,d_2]
  \end{equation}
This implies that the codes $C^{+K} = [n,k_1,d_1]$ and
$C^{\perp} = [n,k_2,d_2]$ have sizes $k_1$, $k_2$ related by
  \begin{equation}
k_1 + k_2 = n + K  \label{k1k2}
  \end{equation}
Since all codes satisfy the Hamming bound (\ref{Hammn}), both $C^{+K}$ and
$C^{\perp}$ do so. Substituting in equation (\ref{k1k2}), this implies
  \begin{equation}
 \frac{K}{n} \le 1 -  H\left(\frac{d_1}{2n}\right)
                      -  H\left(\frac{d_2}{2n}\right) 
\label{limH}  \end{equation}
where the factors $(1 - \zeta)$ have been dropped for clarity (this will
not affect the argument).

Now, provided the parameters $[n,k_1,d_1]$ satisfy the Gilbert-Varshamov 
(G-V) bound (\ref{GVn}), then it is certainly possible to find a code 
$C^{+K}$ having size $k_1$ and minimum distance $d_1$. What is the 
condition that such a code will have associated with it a $K$'th order 
subcode $C$ whose dual $C^{\perp}$ has minimum distance $d_2$? I conjecture 
that it is sufficient that $C^{\perp}$ also satisfy the Gilbert-Varshamov 
bound. I have not been able to prove this, but the conjecture seems 
reasonable since it is known that there is an infinite series of self-dual 
codes which satisfy (\ref{GVn}). Therefore in the set $\{ C^{+K} 
\leftrightarrow C \leftrightarrow C^{\perp} \}$ $= \{ [n,n/2+K,d_1] 
\leftrightarrow [n,n/2,d_2] \leftrightarrow [n,n/2,d_2] \} $, both $C$ and 
$C^{\perp}$ can satisfy the G-V bound simultaneously. In passing from $C$ 
to $C^{+K}$ in this case, one does not expect the minimum distance to fall 
especially rapidly, so it is reasonable to suppose that $C^{+K}$ can also 
be found satisfying the G-V bound.

{\em Nota bene} since submitting this
manuscript I have learnt that Calderbank and Shor 1996 have proved the
above conjecture for the case $d_1 = d_2$, by proving that 
the G-V bound is a sufficient condition for the existence of a
{\em weakly} self-dual code, ie one containing its dual. These authors
have reported an independent derivation of the most important
result (theorems 5 and 6 combined) of the present work.

It will be assumed, then, that a sufficient condition for $K$ qubits to be 
encoded with minimum distances $d_1$, $d_2$, is that $C^{+K}$ and 
$C^{\perp}$ both satisfy the Gilbert-Varshamov bound. Substituting this 
bound (\ref{GVn}) in equation (\ref{k1k2}) leads to 
  \begin{equation}
    \frac{K}{n} \ge 1 -  H\left(\frac{d_1}{n}\right)
                      -  H\left(\frac{d_2}{n}\right) 
\label{limGV}  \end{equation}
Inequalities (\ref{limH}) and (\ref{limGV}) are closely related to 
Shannon's main theorem in the classical regime. The classical regime 
corresponds to the limit $d_2/n \rightarrow 0$, $d_1 = d$, in which case we 
obtain $1- H(d/2n) \ge K/n \ge 1 - H(d/n)$. The context in which we have 
been working throughout corresponds classically to a binary symmetric 
channel, having capacity $C(p) = 1 - H(p)$ where $p$ is the error 
probability. Shannon's theorem states that the rate $K/n$ can be 
arbitrarily close to capacity, while allowing error-free transmission. This 
implies $K/n \sim 1 - H(p)$ is possible for a code of average distance 
$\overline{d} = 2 np(1 + \zeta)$ with $\zeta$ arbitrarily close to zero. 
The averaging employed here involves various technicalities which are 
discussed in standard texts; a good introduction is given by Hamming 1986. 
For our present purposes, we note simply that Shannon's theorem gives $K/n 
\sim 1 - H(\overline{d}/2n)$. Comparing this with inequality (\ref{limH}), 
one sees that classically the Hamming bound gives a good guide to the 
limits of what is possible for an average distance between codewords, even 
though the {\em minimum} distance cannot not reach the upper limit of the 
Hamming bound, and indeed more restrictive bounds are known (see 
MacWilliams \& Sloane 1977). This suggests that the Hamming bound is a 
useful indicator in general, ie that codes which `approach' it in an 
average way do exist. 

Returning to the quantum regime, let us consider for simplicity the case 
$d_1 = d_2 = d$. This is the type of coding one would choose if the 
probabilities of errors in bases 1 and 2 were equal. If they are not equal, 
one can always choose $d$ sufficiently large to allow correction in the 
most error-prone basis, then it will also be more than sufficient for 
correction in the other basis. For $d_1 = d_2 = d$, inequalities 
(\ref{limH}) and (\ref{limGV}) give 
  \begin{equation}
H\left(\frac{d}{n}\right) \ge \frac{1}{2} \left(1 - \frac{K}{n} \right)
\ge H\left(\frac{d}{2n}\right),
  \label{Knlim}  \end{equation}
which, in the case $n \gg K$, implies
  \begin{equation}
H^{-1} \left( \frac{1}{2} \right) \le \frac{d}{n}
\le 2 H^{-1} \left( \frac{1}{2} \right).
  \label{dnlimit}  \end{equation}
The inverse entropy function $H^{-1}(x)$ is defined for $0 < x \le 1/2$ by 
$H^{-1}(x) = y$ iff $x = H(y)$. Using $H^{-1}(1/2) \simeq 0.110028$, we
find that encoding $K$ qubits using $n \gg K$ 
allows $d/n$ to be greater than $0.11$, while $d/n$ is certainly less than 
$0.22006$. These limits are shown on figure \ref{Hplot}. 

At this point, a complete discussion would introduce the notion of the 
capacity of a noisy quantum channel. The capacity would be limited by error 
rates, and one would investigate whether error-free transmission is 
possible at rates close to capacity, as in Shannon's theorem. However, the 
capacity of a noisy quantum channel is not yet understood, and the present 
author has not developed a satisfactory definition. The equivalent of 
Shannon's noiseless coding theorem has been developed for the quantum 
regime by Schumacher 1995, and it is found that the number $K$ of qubits is 
a useful measure of ``amount of information'' in the quantum case, as one 
would hope. To understand the effect of noise (ie errors) in the quantum 
regime, I conjecture that it is useful to model errors in a way analagous 
to that employed in classical information theory. That is, we {\em assume} 
that decoherence, relaxation and so on in a real
quantum computer or information channel can be {\em modelled} by 
a stochastic treatment, in which, between two defined times, each qubit
either defects (undergoes an arbitrary error), or follows the error-free 
evolution governed by the known parts of the system Hamiltonian. Which of 
these two occurs for any given qubit during any given time interval is a 
random decision, the defection occuring with probability $p$. This model is 
somewhat akin to the `quantum jump' or `quantum Monte Carlo' models of 
dissipative processes in quantum mechanics introduced by several authors in 
different contexts (Carmichael 1991; Dalibard {\em et al.} 1992; Dum {\em 
et al.} 1992; Gisin 1984; M\o lmer {\em et al.} 1993 and references 
therein). This similarity suggests that the model can provide a realistic 
description of a large class of real error processes\footnote{Note,
however, that the occurence of a random defection is not to be 
identified exactly with a quantum jump occuring in the quantum Monte Carlo 
method of Dalibard {\em et al.} 1992, M\o lmer {\em et al.} 1993, since in 
that method an error term still appears in the state when no quantum jump 
occurs, due to the non-Hermitian part of the Hamiltonian employed during 
the periods of evolution between jumps. The quantum Monte Carlo method has
exactly the stochastic form we require when only phase relaxation occurs in
one basis (``relaxation of type $T_2$'', cf section \ref{sec:spa}).}. 
In fact the assumption we make is not that error processes can be {\em 
fully} modelled in this way, but merely that a scheme which can correct 
this assumed type of error will be able to correct satisfactorily the 
errors in a real physical computer.

With our stochastic model of errors, an argument using the law of large 
numbers can be employed to show that the probability that uncorrectable 
errors occur falls to zero when codes of long enough minimum distance are 
employed, just as in classical information theory. Thus, in the stochastic 
model, the probability that exactly $x$ errors (defections) occur
among $n$ qubits is 
given by the binomial distribution, if we assume that
errors in different qubits are independent.
The probability that any number up to $x$ qubits defect is 
  \begin{equation}
F(x) = \sum_{i=0}^{x}
\left( \raisebox{-0.5ex}{$\stackrel{n}{\scriptstyle i}$} \right)
p^i (1-p)^{n-i}.
  \end{equation}
where $p$ is the probability of defection of a single qubit, during some 
defined interval of time $t \rightarrow t + \Delta t$.
When $n, n p \gg 1$, this can be approximated using the error function: 
$F(x) \simeq {\rm erf}( (x - \mu)/ \sigma \sqrt{2} )$
where $\mu = n p$, $\sigma = \sqrt{ n p (1-p) }$.
If an $x$-error correction scheme is implemented (using a code of distance 
$d=2 x + 1$), then $F(x)$ is the probability that the code can be corrected 
successfully. When the correction is successful, the state of the quantum 
computer is {\em exactly} what it should be (the assumptions of the 
stochastic model permit this `unphysical' conclusion). 

If the whole computation requires a total number $T$ of time steps, each of 
duration $\Delta t$, and error correction is carried out at the end of each 
time step, then the probability that the whole computation is free of 
errors is 
  \begin{equation}
P(n,p,d,T) = \left( F\left(x = \left\lfloor (d-1)/2 \right\rfloor
\right) \right)^T
  \label{PT} \end{equation}
In the case of the binomial distribution, the law of large numbers is 
expressed by the fact that once the number of correctable errors $x = 
\lfloor (d-1)/2 \rfloor$ becomes larger than the mean number of errors $\mu 
= n p$, the error function $F(x)$ becomes arbitrarily close to 1 as $n$ is 
increased. To see just how close, we use the asymptotic expansion
$1 - {\rm erf}(z) \simeq (\exp (-z^2)/2\sqrt{\pi})(1-1/2 z^2 + \cdots)$,
for $z \gg 1$, which gives
  \begin{eqnarray*}
&& P(\mbox{recover successfully after one timestep}) \simeq F(x=d/2) \\
&\simeq& 1 - \frac{1}{d/2n - p} \sqrt{ \frac{ 2p(1-p)}{n\pi} \,} \,
\exp \left( \frac{ -n (d/2n - p)^2 }{ 2p(1-p) }\right) \\
\mbox{where} && \frac{d}{2n} \sim H^{-1}((1-K/n)/2) 
  \end{eqnarray*}
using inequalities (\ref{Knlim}). The probability that the
computer cannot recover from errors falls exponentially with $n$, as
long as codes are used which keep close to the maximum possible correction
ability (ie Hamming distance), and as long as $p$ is below an upper
bound which does not depend on $n$ or $K$. It has been supposed that
such exponential stabilisation would be impossible for a quantum computer.
Indeed, it is a surprising result which goes right against the usual
conclusion of the Schr\"odinger cat paradox, in which macroscopic
superpositions appear to be inherently unstable, and unstabilisable. 
However, using inequalities 
(\ref{dnlimit}) (cf figure \ref{Hplot}), we find that error-free 
computation is guaranteed to be be possible if $p < \sim 0.055$.
Error-free computation is impossible if $p > \sim 0.11003$ if
correction is attempted using the type of coding method derived
in this paper. However, we
have not ruled out the possible existence of more powerful correction
methods which would allow this upper limit on $p$ to be increased.

As an example, consider a computer requiring $K=1000$ qubits, which are 
encoded using a set of $n=10000$ qubits. Inequalities (\ref{Knlim}) allow 
$d_1 = d_2 = 939$. Suppose the error (ie defection) probability
during each time step is 
$p = 0.04$, and $T=10000$ time steps are required. During each step, on 
average $400$ errors occur, and the standard deviation of the error 
distribution is about $20$. The probability of error-free computation is, 
from equation (\ref{PT}), $P(n,p,d,T) \simeq 0.01$. If $p$ is reduced to 
$p=0.03$, on the other hand, then $F(x) \simeq 1 - 4 \times 10^{-23}$ and 
error-free computation is almost certain for any reasonable length $T$ of 
the calculation. 

\section{Concluding remarks}

The error correction methods which have been presented involve many 
two-qubit {\sc cnot} operations each time correction is carried out, and 
the whole process only works if these operations can be performed without 
introducing too many extra errors. The number of $1$'s in the parity check 
matrix of a $[n,k,d]$ code is about $k d$, since each message bit must be 
associated with at least $d-1$ parity checks. Therefore, the error 
corrector of such a code involves of the order of $k d$ two-qubit 
operations. For the case $d_1 = d_2$ one uses coding with $k_1 = k_2 \simeq 
n/2$, and correction is carried out in both bases. Therefore the total 
number of operations for one complete correction is of order $n d \simeq 
2 n^2 p$. A logical choice would be to correct the whole computer every 
time an elementary computing operation is performed. Therefore,
the introduction of error correction causes the total 
number of two-qubit operations to be multiplied by $2 n^2 p$. This is a 
modest extension of the resources necessary to complete a computation, 
since $n$ itself does not increase faster than the number $K$ of bits of 
quantum information employed. The great gains in computing power associated 
with ``quantum parallelism'' are retained in the corrected noisy computer.

It may come as a surprise that the condition for error-free computation 
derived above is simply an upper bound $p < H^{-1}(1/2) \simeq 0.11$ on the 
error probability, rather than a scaling law for 
$n$ as a function of $K$ and $p$.
However, this is in the nature of the approach we have adopted, since when 
$n p \gg 1$, the number of errors that actually occur during any one time 
step is almost certainly very close to the average number $n p$, so the 
whole battle is won or lost on the ability to correct this number of 
errors. The corresponding limit in the classical regime is $p < H^{-1}(1) = 
1/2$, above which error-free communication is impossible and the channel 
capacity falls to zero. Indeed, if the only errors that occur in the 
quantum case are phase errors in one of the bases (say basis 1), then we 
can afford to use $d_1 = 1$, and the coding problem reduces to the 
classical one, so $p$ can approach $H^{-1}(1)$ once again. The factor $1/2$ 
rather than $1$ appearing in the inverse entropy function arose because it 
was assumed just before inequalities (\ref{Knlim}) that arbitrary unknown 
errors will require correction in both basis 1 and basis 2. It is
here that 
the difference between a qubit and a classical bit enters: the extra 
degrees of freedom associated with the qubit mean that we do not know, in 
general, in which basis to correct it, so we are forced to correct it in 
two mutually rotated bases. By theorem 3, this means that both a code and 
its dual must be capable of correcting the expected error rate, so both 
$k/n$ and $1-k/n$ are limited by Shannon's theorem. The classical and 
quantum cases can be compared thus: \pagebreak[0]
{\samepage  \begin{eqnarray*} 
{\rm Classical} \rule{6ex}{0cm} &\rule{1cm}{0cm}&
\rule{6ex}{0cm} {\rm Quantum} \\ 
\left. \begin{array}{rcl} 
k/n &<& 1 - H(p) \\
k/n &>& 0 \end{array} \right\} && 
\left. \begin{array}{rcl} 
k/n &<& 1 - H(p) \\
1-k/n &<& 1 - H(p) \end{array} \right\} \; \left( n \gg K\right) \\
\Rightarrow \;\; \;\;H(p) < 1 \;\;\;\; && \Rightarrow \;\;\;\; H(p) < 1/2
  \end{eqnarray*}  }
It should be stressed that the left hand side of this comparison represents 
a well-founded body of knowledge, while the right hand side involves some 
assumptions which remain to be investigated further. 

The whole argument has assumed that the process of error correction does
not itself introduce defection (ie decoherence etc). However
this is an unrealistic assumption,
since the error correction procedure is itself a special kind of quantum
computation. Clearly, the probability of error
during one time step must be reckoned to increase with the number of
operations needed to implement error correction. However, such errors
can be corrected during the next time step, provided that they
affect sufficiently few qubits. The analysis of this in detail
is an important avenue for future work.

In conclusion, the main contribution of this paper, and of
Calderbank and Shor 1996, has been to show how to 
adapt the classical methods of error correction to the quantum regime. 
Theorems 5 and 6 are central. The sections leading up to them introduced the 
ideas, and those following examined the implications. The theorems show
that a macroscopic quantum system can be stabilised by a judicious use 
of unitary operations and dissipative measurements.
This is a type of feedback loop or `quantum servo-control'.
Among the results 
gained along the way are a useful taxonomy of types of multiple-particle 
(or multiple-qubit) interference, and a general insight into how to perform 
quantum purity amplification. These are basic properties of quantum 
theory, the former showing how information can be embodied in many qubits
simultaneously, and the latter showing how quantum communication can
be isolated from noise and evesdropping.
The linear codes we have discussed constitute a 
generalisation of the `Bell basis' to many qubits. 

The obvious need now is for a fuller understanding of the capacity of a 
noisy quantum channel. In particular, it would be useful to find out 
whether a stochastic model for errors in a quantum channel is 
sufficient to enable the error rate for many qubits to be estimated. Also,
the effect of noise during the error correction
process needs to be investigated. 
On the experimental side, an implementation of the simplest 
error correction schemes using 3 or 7 qubits would be a significant step 
forward. 

I thank members of the Oxford Quantum Information group for commenting on 
the manuscript. The author is supported by the Royal Society.

\newpage

\begin{figure}
\caption{Encoding for simplest error correction scheme. Initially
$Q$ is the qubit to be encoded, and the three `encoding' qubits are in
the state $\protect\bet{000}$. Symbols: $\updownarrow =$ state swapping;
$^| \hspace{-1.2ex} \times =$ controlled not in basis 2.}
\label{fig:enc01}
\end{figure}

\begin{figure}
\caption{Simplest error correction scheme. All operations take place
in basis 2. After two {\sc cnot} operations, the lower two qubits
are measured in basis 2. The results are fed to a classical `box'
which then complements ({\sc not} operation) one or more of the
qubits, depending on the measurement results. The two
final {\sc cnot}'s reencode the state (see text).}
\label{fig:cor01}
\end{figure}

\begin{figure}
\caption{Encoder for the simplest scheme enabling single error correction 
in both bases. The multiple {\sc cnot} symbols mean successive {\sc cnot} 
operations carried out between the single control qubit and each of the 
target qubits. The initial state is $\ket{00Q0000}$. The first two {\sc 
cnot} operations prepare for the generation of a superposition of the 
simplex code and its complement. The rest generates the code from this 
preparatory state. The symbol $\bigcirc \hspace{-1.8ex} \scriptstyle R$ 
means the rotation $\ket{0} \rightarrow \protect\bet{0}$, $\ket{1}
\rightarrow \protect\bet{1}$.} 
\label{fig:enc11}
\end{figure}

\begin{figure}
\caption{Error corrector for the code generated by figure 
\protect\ref{fig:enc11}. Multiple {\sc cnot} operations perform parity 
checks. The lower three qubits are measured, and the results used to 
determine which qubits undergo a {\sc not} operation. The scheme is first 
applied in basis 1, then in basis 2 (see text).} 
\label{fig:cor11}
\end{figure}

\begin{figure}
\caption{Alternative method of error correction. Codes need not be corrected
`in place' using the qubits of the computer itself (as in the previous
figures). It may be more convenient to establish the error syndrome using a
set of ancillary qubits. The example shown here carries out the same correction
as the corrector of figure \protect\ref{fig:cor11}.}
\label{fig:anc}
\end{figure}

\begin{figure}
\caption{Asymptotic bounds, in the limit of large $n$, on the rate of
a quantum code. Full curves: inequalities (\protect\ref{Knlim}); dashed
curve: Hamming bound for a classical code (ie quantum case with
phase errors only). The upper full curve is an upper (Hamming) bound
for the codes discussed in the text, the lower full curve is a sufficient
(Gilbert-Varshamov) condition for the existence of the codes discussed.}
\label{Hplot}
\end{figure}

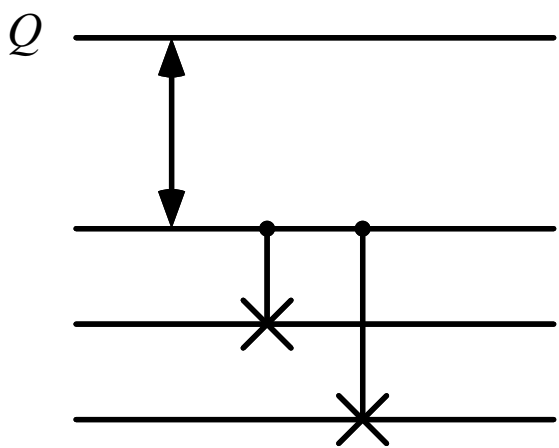
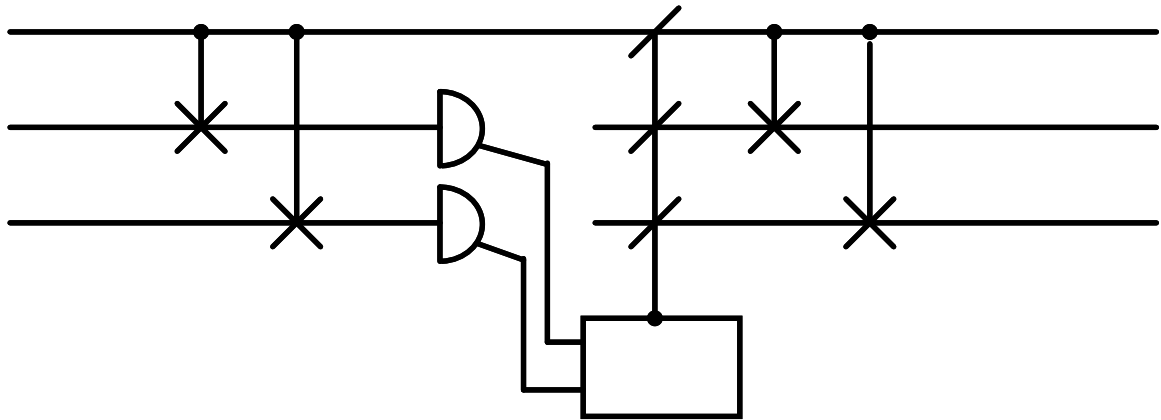
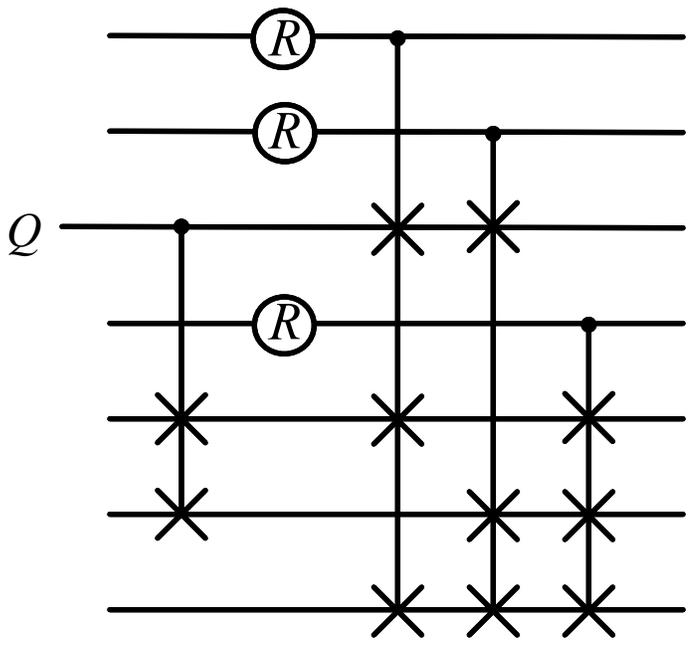
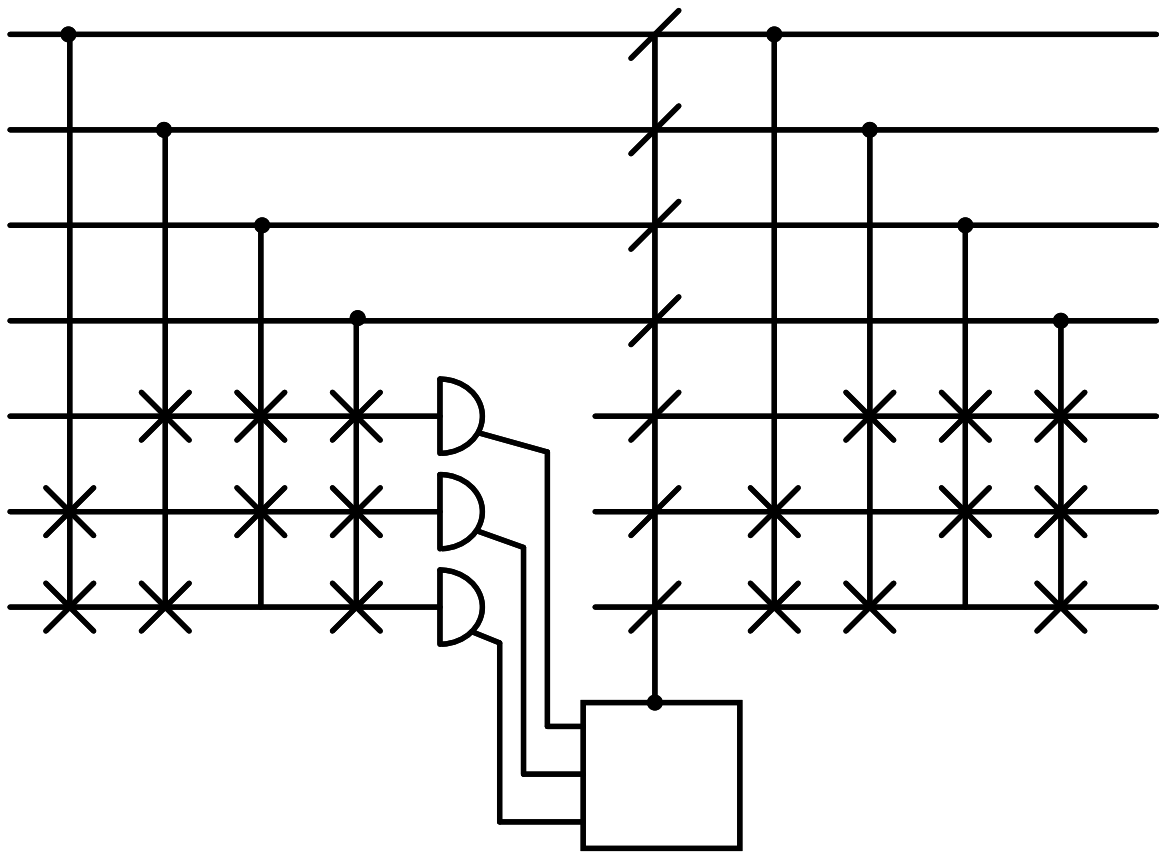
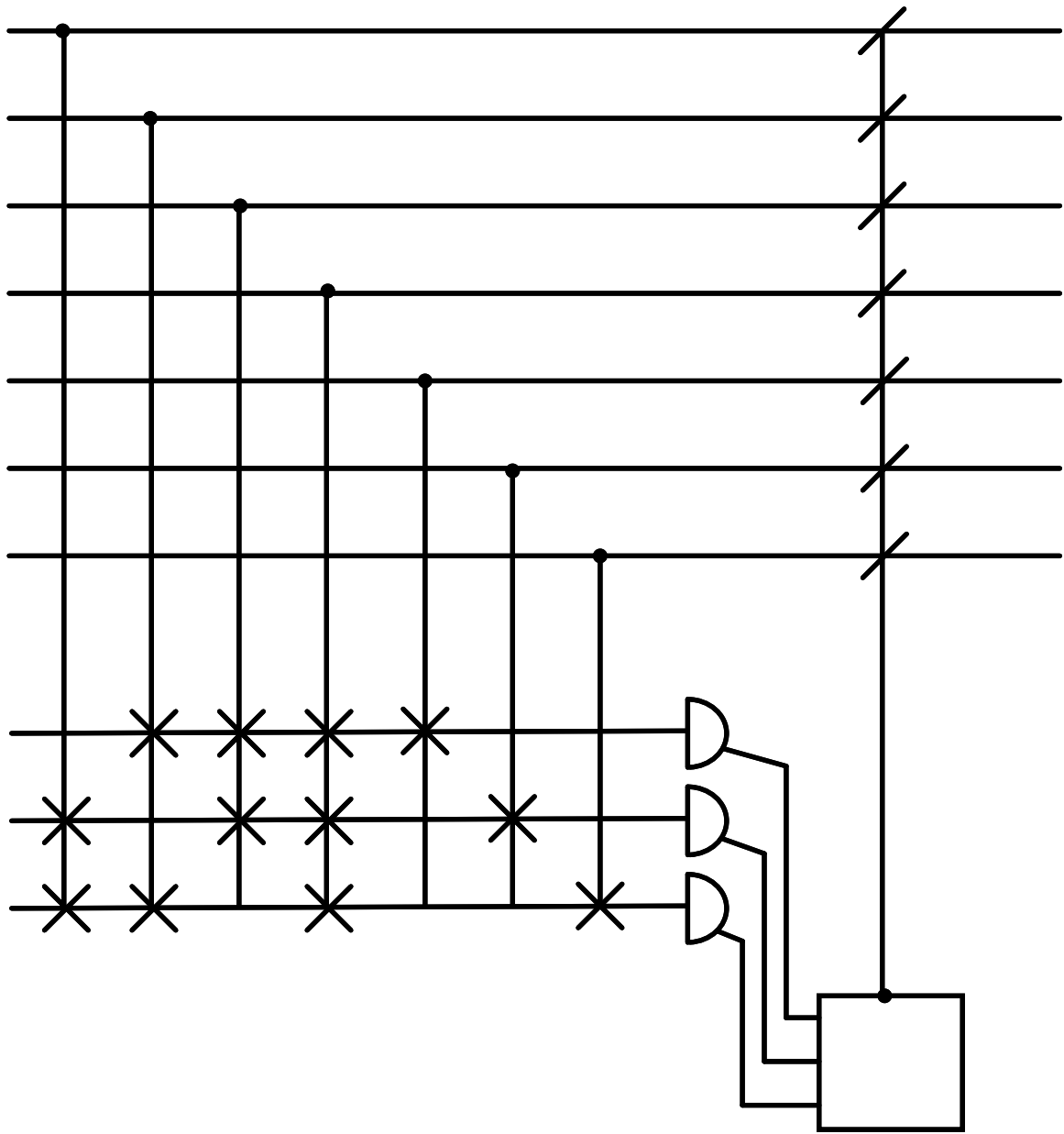
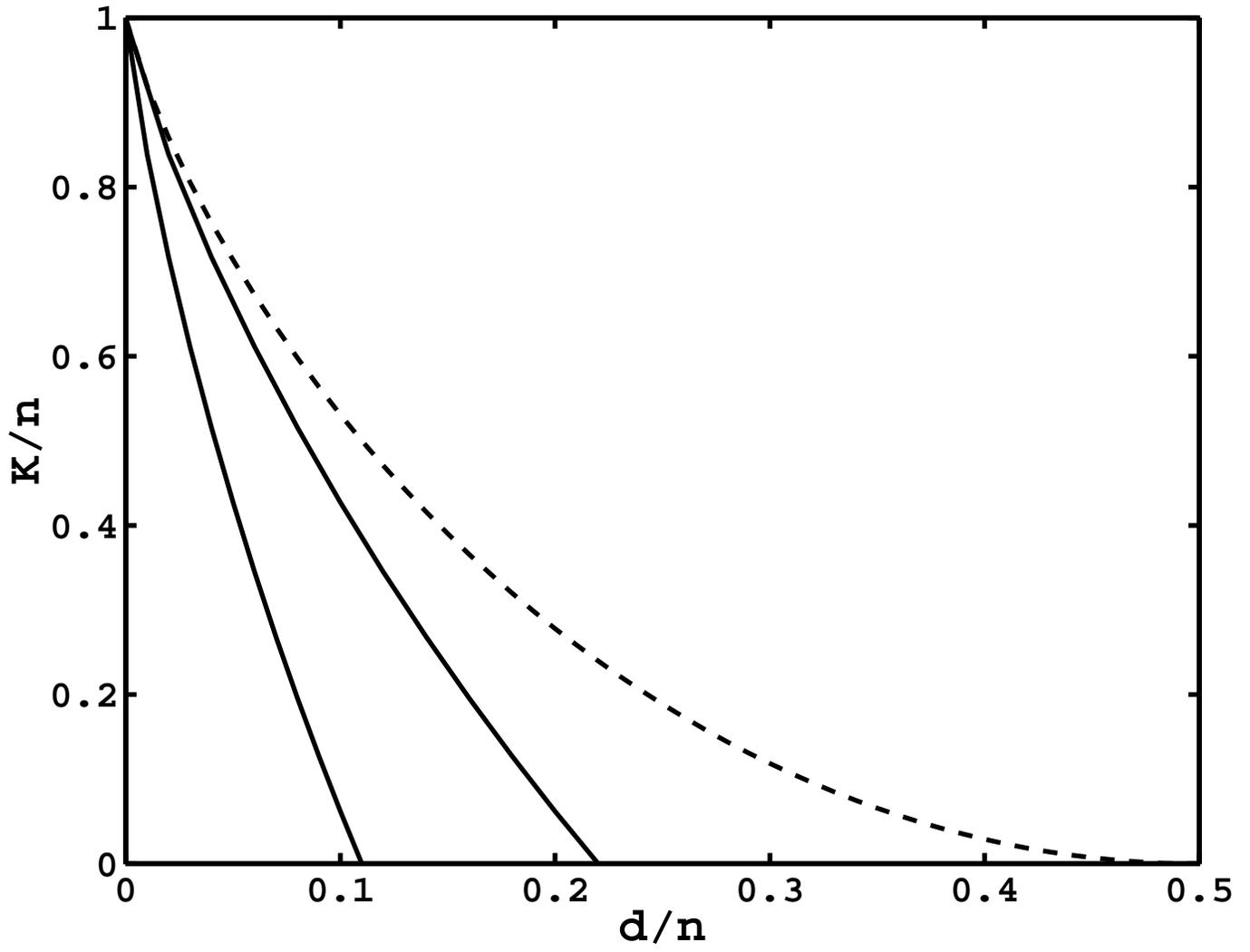
\end{document}